# OBSERVING AND RECOMMENDING FROM A SOCIAL WEB WITH BIASES

## PART I

Web Science Institute (WSI)
PUMP-PRIMING PROJECT

**University of Southampton**

26 January — 11 March 2016

V2.2 last modified by LC on 11/03/2016

**Professor Steffen Staab**
PRINCIPAL INVESTIGATOR
S.R.Staab@soton.ac.uk
Web and Internet Science,
Faculty of Physical Sciences and Engineering

**Dr Sophie Stalla-Bourdillon**
CO-PRINCIPAL INVESTIGATOR
S.Stalla-Bourdillon@soton.ac.uk
Institute for the Law and the Web (iLaws),
Faculty of Business, Law and Art

**Dr Laura Carmichael** (née German)
POST-DOCTORAL RESEARCH FELLOW
leg1r13@soton.ac.uk
Institute for the Law and the Web (iLaws),
Faculty of Business, Law and Art

## Table of Contents

**Acronyms**

DADM = Discrimination-aware data mining

DPD = Data Protection Directive

DPDM = Discrimination prevention in data mining

DSS = Data Support System

EHRC = Equality and Human Rights Commission

FTC = Federal Trade Commission (USA)

GDPR = General Data Protection Regulation

ICO = Information Commissioner's Office (UK)

ILAWS = Institute for the Law and the Web

WSI = Web Science Institute

# 1. INTRODUCTION

## 1.1 Research question

**The research question this project addresses is: how, and to what extent, those directly involved with the design, development and employment of a specific "black box" algorithm can be certain that it is not unlawfully discriminating (directly and/or indirectly) against particular persons with protected characteristics (e.g. gender, race and ethnicity)?**

## 1.2 Automatic socially-sensitive decision-making

Socially-sensitive decision-making is becoming increasingly automated. Automatic decision support and recommendation systems are utilised across a number of industries, e.g. insurance provision, mortgage lending, recruitment, product marketing, financial services and credit [1, p. 643]. For instance, such algorithms may be used to match a particular insurance tariff to *person x*, or assess whether *person y* is suited to a specific vacant employment position.[1] Toshihiro Kamishima et al. [1, p. 643] provide two key reasons for the greater reliance on such automatic decision-making processes: (1) there is an unprecedented amount of (personal) data available from multiple sources (e.g. "demographic information, financial transactions, communication logs, [and] tax payments [1, p. 643]); and (2) the ease of access to "off-the-shelf mining tools".

### 1.2.1 Algorithms can discriminate

Data analytics aims to reveal useful patterns and trends across (often) large sets of data. This increased level of analysis may provide greater opportunities to advance understanding in a number of areas, including medical research and transportation [2, p. 2]. Despite the benefits, there is also a risk that such patterns and trends may be (un)intentionally misused e.g. to unlawfully discriminate against persons with protected characteristics [2, p. 2], [3, p. 1734], [4, p. 4].[2] Moreover, such patterns and trends may be incorrect where they are based on inaccurate data [2, p. 2]. Solon Barocas and Andrew D. Selbst [5, p. 1] further highlight the potential for algorithms to discriminate:[3]

---

[1] For further background information on how complex algorithms are used for daily decision making see: Kevin Slavin's TED Talk: How algorithms shape our world [124]; and [25]. For more information on how businesses utilise data mining see [111]. For some examples of how data analytics are being used see [2, pp. 6-8].

[2] Bart Custers [4, p. 4] states: *"As databases contain large amounts of data, they are increasingly analyzed in automated ways. Among others, data mining technology is applied to statistically determine patterns and trends in large sets of data. The patterns and trends, however, may be easily abused, as they often lead to unwanted or unjustified selection. This may result in the discrimination of particular groups."* Sara Haijan et al. [3, p. 1734] further maintain: *"Data mining is an increasingly important set of analytical processes that allow extracting useful knowledge hidden in large collections of data, especially human and social data sensed by the ubiquitous technologies that support most human activities in our age. As a matter of fact, the new opportunities to extract knowledge and understand human and social complex phenomena increase hand in hand with the risks of violation of fundamental human rights, such as privacy and non-discrimination."* For further background information also see [130].

[3] There is sometimes a misconception that algorithms and "big data" cannot discriminate; this notion has been refuted by a number of commentators. For instance, Kate Crawford – a researcher at Microsoft Research – sought to dispel six big data myths during the DataEDGE Conference 2013 held at the UC Berkeley School of Information [80]. These big data myths (as reported in the New York Times [80]) are: (1) "big data is new"; (2) "big data is objective"; (3) "big data doesn't discriminate"; (4) "big data makes cities smart"; (5) "big data is anonymous"; and (6) "you can opt out". Ann Bevitt [90, p. 7] further states: *"It is often assumed that algorithms and automated processes are neutral and unbiased, but this is not necessarily the case: big data sometimes has the effect of reinforcing existing stereotyping and prejudices."*

> "Advocates of algorithmic techniques like data mining argue that they eliminate human biases from the decision-making process. But an algorithm is only as good as the data it works with. Data mining can inherit the prejudices of prior decision-makers or reflect the widespread biases that persist in society at large. Often, the "patterns" it discovers are simply preexisting societal patterns of inequality and exclusion. Unthinking reliance on data mining can deny members of vulnerable groups full participation in society. Worse still, because the resulting discrimination is almost always an unintentional emergent property of the algorithm's use rather than a conscious choice by its programmers, it can be unusually hard to identify the source of the problem or to explain it to a court."

In consequence, digital profiling can lead to unjust stereotyping, unfair discrimination based on price and information asymmetries.[4] While concerns over potential inequalities arising within the data analytics environment is not new (e.g. [6] and [7]), a recent US Federal Trade Commission (FTC) report published in January 2016 – "Big Data: A Tool for Inclusion or Exclusion" [2] – re-affirms these apprehensions.

A key example of such inequality and exclusion within the data analytics environment is "weblining" [6], [7, p. 382] – i.e. where online service providers utilise "profiling"[5] in order to offer and/or deny particular services, products and/or information to certain groups of individuals.[6] It is often the case that to remain legally-compliant, automated decision-making systems avoid the use of protected characteristics in their analyses (e.g. the UK Equality Act 2010, section 4 defines nine legally protected characteristics, including: age, race and sexual orientation). However, the key issue here is that other background data (e.g. neighbourhood) can act as proxies for these protected characteristics (e.g. ethnicity).[7] For instance, Lori Andrews [8] reports that stereotyping people with similar likes and dislikes could have a bearing on credit scores. For example, if the majority of people who frequently shop at *retailer y* have a poor payment history, a lender may take this aggregate data into account for all its (potential) customers [8]. Therefore, if *person x* often shops at *retailer y* and has a good credit payment history, their credit score could be lowered due to the aggregate data despite their good individual payment history [8].

### 1.2.2 Example i – medical school admission

The following example of automated discrimination not only shows that this is not a new phenomenon, but that an automated admissions procedure may be unconsciously, and therefore unintentionally, discriminatory.

---

[4] For further background information about ethics and web data mining see [128]; for further background about information asymmetries, big data and the insurance market see [131].

[5] Lee A. Bygrave [122, p. 17] provides a definition of "profiling": *"Generally speaking, profiling is the process of inferring a set of characteristics (typically behavioural) about an individual person or collective entity and then treating that person/entity (or other persons/entities) in the light of these characteristics. As such, the profiling process has two main components: (i) profile generation – the process of inferring a profile; (ii) profile application – the process of treating persons/entities in light of this profile."*

[6] Edward C. Baig et al. [103] define weblining as follows: *"Some companies use the gold mine of consumer data to discriminate against customers who don't make the grade. You might call it "Weblining.""* Marcia Stepanek [6] states: *"If you do get Weblined, you may not even know it. For that, thank the wonders of "personalization" software, which quietly analyzes data and dishes up products or services seemingly tailored to you. On the surface, it might seem as if you are getting something really spiffy--a suggestion for a vanity checkbook, or a reminder of your daughter's birthday. In fact, somebody--or perhaps just a computer--has decided what you are fit to see, sample, or buy, based on sometimes-crude calculations of what you are worth to the firm. "Companies can segment without being very obvious about it," says Forrester's Chatham. "But just because it's personalized doesn't mean that what you're seeing and getting is first-class.""*

[7] For further information, Rick Swedloff [132] explores the use of opaque proxies in the context of risk classification within insurance.

The case of St. George's Hospital Medical School provides a real-world example of unconscious bias where the computer program designed to screen its applicants, unknowingly but, unfairly discriminated against applicants who were female and/or from ethnic minorities.[8]

During the 1970s and 1980s, St. George's Hospital Medical School used a computer program for initial screening of applicants; by 1982 all initial screening was automated [9, p. 657]. The program used information from applicants' University Central Council for Admission (UCCA) forms, which contained no reference to ethnicity. However, the computer program deduced the ethnicity of candidates from surnames and places of birth. The program was found to unfairly discriminate against female applicants and those from ethnic minorities, as there were less likely to be selected for interview [9, p. 657].

The key point highlighted by Stella Lowry and Gordon Macpherson [9, p. 657] is that: "the program was not introducing new bias but merely reflecting that already in the system." In 1988, while the UK Commission for Racial Inequality did not serve a non-discrimination notice on St. George's Hospital Medical School, it did find the school guilty of racial and sexual discrimination in its admission policy [9, p. 657].

### 1.2.3 Example ii – predictive policing

The following example demonstrates how predictive algorithms – such as PredPol [10] – are assisting police forces by highlighting crime hotspots.

The USA's National Institute of Justice (NIJ) [11] defines predictive policing:

*"Predictive policing tries to harness the power of information, geospatial technologies and evidence-based intervention models to reduce crime and improve public safety. This two-pronged approach — applying advanced analytics to various data sets, in conjunction with intervention models — can move law enforcement from reacting to crimes into the realm of predicting what and where something is likely to happen and deploying resources accordingly."*

A number of police force use predictive policing, including the Chicago Police Department (USA) [12], the Los Angeles Police Department (USA) [10], and Kent Police (UK) [13]. However, predictive policing has attracted some controversy. For instance, the data used might contain (historic) biases against certain groups of individuals – leading Jeremy Kun [12] to ask: *"But do these algorithms discriminate, treating low-income and black neighborhoods and their inhabitants unfairly?"*

### 1.2.4 Example iii – retail: pregnancy prediction

The case of Target's pregnancy-prediction offers a real-world example of how shopping patterns can be used to predict pregnancy amongst a customer-base to help improve targeted advertising. This case was originally reported by Charles Duhigg [14] in *The New York Times Magazine* during February 2012.

As is the case with many brands, the US retailer Target collects data on its customers by assigning each person a unique code: guest ID number, recording their purchases and demographics information (e.g. age, marital status and address) [14]. All these data can be used for predictive analysis in order to better understand (individual) shopping habits; brands are then better-equipped to recommend products and offer more appropriate vouchers [14]. Shopping habits are largely habitual, however there are major events that can disrupt these habits e.g. pregnancy. Therefore, Target wanted to create a pregnancy-prediction model in order to obtain further custom.

---

[8] Barocas and Selbst [5, p. 12] also reference this example.

Pole began by analysing Target's Baby Shower Registry to find shopping patterns of expectant mothers. He discovered 25 products that could be used for a pregnancy prediction score e.g. unscented lotion. Target was able to predict who of its customer-base were most likely to be pregnant and deliver targeted advertising e.g. vouchers for baby products.

(Note (1) that the UK Information Commissioner's Office (ICO) [15, p. 15] also raises this example as part of its big data and data protection guidance. ICO makes clear that: *"Fairness involves a wider assessment of whether the processing is within the reasonable expectations of the individuals concerned. For example, every aspect of analysing loyalty card data to improve marketing should not always automatically be considered fair, or within customer expectations."*

Note (2) that Kaiser Fung [16] questions whether this account reported in the *New York Times* takes into account false positives i.e. the number of people who are not pregnant who still received coupons for baby products.[9])

### 1.2.5 Lack of awareness among data subjects

The use of profiling may be more obvious and beneficial to end users in certain situations (e.g. an online music store recommending new music based on personal preferences). However, in many other circumstances the end user may not be aware that: (1) automated decision-making processes are taking place [6]; and (2) the proportion of such decisions which are potentially leading to unjust treatment and censorship. Eli Pariser's TED Talk: Beware online "filter bubbles" [17] draws attention to how online personalisation often occurs without the direct knowledge of the end user. For instance, a web user may assume that the content displayed on a particular website is identical for all web users visiting that page at the same time, where in reality the content is automatically generated according with the perceived preferences of that unique web user. Moreover, end users may not realise or appreciate the fact that web providers are not neutral entities and (as with all institutions) have their own set of biases.[10]

In addition, web users may not be completely aware of the types of and ease with which sensitive personal information that can be derived from their digital footprints. For instance, Michal Kosinski et al. [18] found that highly-sensitive personal attributes – such as: "sexual orientation, ethnicity, religious and political views, personality traits, intelligence, happiness, use of addictive substances, parental separation, age, and gender" [18, p. 5802] – could be automatically and reliably predicted from users' Facebook Likes. Furthermore, other studies have focused on predicting demographics information, such as: gender in social networks (e.g. [19] and [20]) and from mobile phone data (e.g. [21]); and age in social networks (e.g. [22] and [23]). A key concern is that sensitive personal information (e.g. gender and sexual orientation) can still be inferred even for those most savvy web users who deliberately withhold such information from web providers.

---

[9] This is also referenced by [131, p. 325].

[10] For more information concerning online moral bias see Andreas Ekström's TED Talk – "The moral bias behind your search results" [125]. A recent example of such online moral bias is Google showing anti-radicalisation links to users searching for extremist-related content [92], [91].

### 1.3 Project aims: addressing the "black box" algorithm problem

For those directly involved with the design, development and utilisation of such automated decision-making systems, it is of paramount importance that they remain legally compliant and do not unlawfully discriminate. However, in the case of many algorithms, it is not clear (even for the technical expert) how the automated decision-making system altogether works, i.e. an algorithm appears as a "black box".[11] Oxford English Dictionaries Online [24] defines black box (in terms of computing): "A complex system or device whose internal workings are hidden or not readily understood." Frank Pasquale [25, p. 3] offers a further definition of a black box: "[…] a system whose workings are mysterious; we can observe its inputs and outputs, but we cannot tell how one becomes the other."

It is therefore difficult to ensure that a particular black box algorithm is legally compliant when an in-house technical professional (and legal expert) only have access to its input and outputs, and indefinite understanding of its internal workings. In consequence, this project aims to explore the ways in which those directly involved with the design, development and utilisation of such black box algorithms can be certain – and if possible offer some form of guarantee – that there is no unlawful discrimination taking place.

> **This 30-day project therefore provides a feasibility study, which makes preparations for development of an interdisciplinary approach to automated discrimination, which would include a model for representing norms describing undesirable biases against sensitive data attributes and metrics for describing situations of norm violation and compliance.**

### 1.4 Structure of the report

In order to begin to address the research question, this report first examines the current legal framework that underpins the digital data analytics environment in Section 2 by examining what constitutes unlawful discrimination. This brief legal overview primarily focuses on: (1) the UK Equality Act 2010; and (2) the concept of profiling under the European Data Protection Directive and recently agreed General Data Protection Regulation. Section 3 examines some existing methodologies used for uncovering instances of (in)direct discrimination generated by certain black box algorithms; a particular focus is given to the area of discrimination-aware data mining. Section 4 lays the foundations for the model (which aims to represent norms describing undesirable biases against sensitive data attributes and metrics for describing situations of norm violation and compliance) based on findings from Sections 1-3. Section 5 finalises this report by outlining its key conclusions and areas for future work.

---

[11] For instance, no-one understands exactly how (convoluted) neural networks function yet such algorithms provide good predictions.

## 2. THE LEGAL APPROACH TO AUTOMATED DISCRIMINATION

### 2.1 Introduction

Equality law and data protection law appear to be the most pertinent areas of law when considering automated discrimination. First, equality law defines the parameters of unlawful discrimination, including the characteristics that are legally-protected, such as gender and age. Second, data protection law (under the recently agreed General Data Protection Regulation) regulates profiling, automated decision-making and the use of special categories of personal data, such as genetic information and political opinions.

While other areas of law also have a prominent role in this domain (e.g. insurance law, and intellectual property law – which may prevent third parties having full access to how automated decision systems work), this section specifically focuses on examining some of the key equality and data protection law issues pertaining to automated discrimination.

### 2.2 Equality law – brief overview

The Equality Act 2010 [26] entered into force on 1 October 2010 and provides a single framework for UK discrimination law; it draws together over 116 former legislative instruments, such as the Equal Pay Act 1970, the Sex Discrimination Act 1975, the Race Relations Act 1976, and the Disability Discrimination Act 1995 [27].

> **Defining unlawful discrimination:**
> Direct discrimination is defined by Section 13(1) the UK Equality Act 2010 [26] as: *"A treats B less favourably than A treats or would treat others."*
> Indirect discrimination is defined by Section 19(1) the Equality Act 2010 [26] as: *"A applies to B a provision, criterion or practice which is discriminatory in relation to a relevant protected characteristic of B's."*

The legal framework appears to offer a limited approach to automated discrimination. First, Section 4 of the Equality Act 2010 provides nine categories of protected characteristics:

> **(1) Age, (2) disability, (3) gender reassignment, (4) marriage and civil partnership, (5) pregnancy and maternity, (6) race, (7) religion or belief, (8) sex, and (9) sexual orientation.**

However, this list is restrictive, as automated decision-making algorithms may use sensitive attributes that go beyond this list. For instance, there is no category protecting socio-economic status i.e. persons with low incomes (note that France is considering whether to add poverty as a protected characteristic [28].) Furthermore, genetic information is not a protected characteristic under the Equality Act; however, is does appear as a special category of personal data under the General Data Protection Regulation (see Section 2.3.1).[12]

Sections 158-159 of the Equality Act 2010 also permit positive action under (also known as affirmative action in the USA) under certain circumstances. This is important to note, as some biased black box algorithm outputs may be lawfully discriminating where the principle of positive action is in force.

---

[12] For more information about genetic information in insurance see [133].

### 2.2.1 Example iv – car insurance: gender as an actuarial factor

One of the key European judgments in the area of discrimination and insurance is the case of *Association Belge des Consommateurs Test-Achats ASBL v Conseil des Ministres (C-236/09) [2012] 1 W.L.R. 1933*, where the Belgian Constitutional Court requested a preliminary ruling on whether an insured person's sex could be taken into account as an actuarial factor by private insurers.[13] On 1 March 2011, the European Court of Justice (ECJ) held that insurers should treat men and female customers equally [29]. Insurers were given until 21 December 2012 to alter their pricing policies [29].

However, in 2015, Stephen McDonald delivered a report to the Royal Economic Society concerning (potential) indirect gender-based discrimination [30], [31]. McDonald found that men under 50 years old still appear to pay more for their car insurance than women; due to a loophole where insurers can utilise a person's job title as an actuarial factor [30]. In consequence, those in male-dominated jobs pay more (e.g. civil engineers pay 13% above average on insurance premiums) than those in female-dominated jobs (e.g. dental nurses pay 10% below average on insurance premiums) [30].

### 2.2.2 Public sector equality duty

Section 149 of the Equality Act 2010 confers a public sector duty on: (1) public authorities and (2) non-public authorities who exercise a public function. Section 149(1) states:

> **SECTION 149 OF THE EQUALITY ACT 2010**
> **"149 Public sector equality duty**
> (1) A public authority must, in the exercise of its functions, have due regard to the need to—
> (a) eliminate discrimination, harassment, victimisation and any other conduct that is prohibited by or under this Act;
> (b) advance equality of opportunity between persons who share a relevant protected characteristic and persons who do not share it;
> (c) foster good relations between persons who share a relevant protected characteristic and persons who do not share it."

This public sector duty may involve the use of equality monitoring and equality impact assessments (which are not a legal obligation).[14] For example, the Equality Challenge Unit [32] provides guidance to the UK higher education sector on such matters. Equality monitoring data could be used as a potential tool to further assist with discrimination assessment.

---

[13] For further information about this judgment see [110].

[14] However, there has been criticism that such monitoring goes too far, and this "red-tape" burden needs to be reduced – see [126] for further information.

## 2.3 Data protection law – brief overview

### 2.3.1 From Directive to Regulation

Article 15 of the European Data Protection Directive 95/46/EC and, its recently agreed replacement, Article 20 of the European General Data Protection Regulation afford data subjects a right to be heard in relation to alleged unfair automated decision-making.[15] Note that while the General Data Protection Regulation (GDPR) was approved by the European Parliament, European Commission and European Council [33] on 15 December 2015, the official texts are yet to be published [correct on 19 February 2016].

> **ARTICLE 15 OF THE DATA PROTECTION DIRECTIVE 95/46/EC**
> *"Automated individual decisions*
> *1. Member States shall grant the right to every person not to be subject to a decision which produces legal effects concerning him or significantly affects him and which is based solely on automated processing of data intended to evaluate certain personal aspects relating to him, such as his performance at work, creditworthiness, reliability, conduct, etc.*
>
> *2. Subject to the other Articles of this Directive, Member States shall provide that a person may be subjected to a decision of the kind referred to in paragraph 1 if that decision:*
>
> *(a) is taken in the course of the entering into or performance of a contract, provided the request for the entering into or the performance of the contract, lodged by the data subject, has been satisfied or that there are suitable measures to safeguard his legitimate interests, such as arrangements allowing him to put his point of view; or*
>
> *(b) is authorized by a law which also lays down measures to safeguard the data subject's legitimate interests."* [34]

> **GENERAL DATA PROTECTION REGULATION (European Commission draft text 2012):**
> *"Article 20 Measures based on profiling*
> *1. Every natural person shall have the right not to be subject to a measure which produces legal effects concerning this natural person or significantly affects this natural person, and which is based solely on automated processing intended to evaluate certain personal aspects relating to this natural person or to analyse or predict in particular the natural person's performance at work, economic situation, location, health, personal preferences, reliability or behaviour.*
>
> *2. Subject to the other provisions of this Regulation, a person may be subjected to a measure of the kind referred to in paragraph 1 only if the processing:*
> *(a) is carried out in the course of the entering into, or performance of, a contract, where the request for the entering into or the performance of the contract, lodged by the data subject, has been satisfied or where suitable measures to safeguard the data subject's legitimate interests have been adduced, such as the right to obtain human intervention; or*
> *(b) is expressly authorized by a Union or Member State law which also lays down suitable measures to safeguard the data subject's legitimate interests; or*
> *(c) is based on the data subject's consent, subject to the conditions laid down in Article 7 and to suitable safeguards.*
>
> *3. Automated processing of personal data intended to evaluate certain personal aspects relating to a natural person shall not be based solely on the special categories of personal data referred to in Article 9.*
>
> *4. In the cases referred to in paragraph 2, the information to be provided by the controller under Article 14 shall include information as to the existence of processing for a measure of the kind referred to in paragraph 1 and the envisaged effects of such processing on the data subject.*
>
> *5. The Commission shall be empowered to adopt delegated acts in accordance with Article 86 for the purpose of further specifying the criteria and conditions for suitable measures to safeguard the data subject's legitimate interests referred to in paragraph 2."* [35]

---

[15] An article published in 2001 by Lee A. Bygrave [122] explores to what extent Article 15 of the Data Protection Directive 95/46/EC "may have a meaningful impact on automated profiling practices" [122, p. 17]. Bart W. Schermer also provides a more recent article on the legalities of automated profiling [94]. Furthermore, the Information Commissioner's Office (ICO) [121] offers a couple of examples when such automated decision rights may arise under the Data Protection Directive. In UK law, section 12 of the Data Protection Act currently covers: "Rights in relation to automated decision-taking."

Firstly, while there is no explicit mention of the term "profiling" by the Data Protection Directive 95/46/EC, it is defined by the General Data Protection Regulation [36]. Secondly, the rights of the data subject appear to be slightly more robust in the new Regulation; in as much as there is a right to "Information to the data subject" (Article 14), "Right of access for the data subject" (Article 15), and necessity to conduct data protection assessments (Article 33).[16] However, does the right to information (Article 14) go far enough to give the data subject sufficient means to understand whether there is and/or has been any form of discrimination caused by a specific algorithm? The answer is: most probably not. Therefore, how can a data subject demonstrate that they have been discriminated against without this information?

In addition, pursuant to Article 14a, data subjects have a right to informed about the logic involved in automated decision-making:

> **GENERAL DATA PROTECTION REGULATION (2015 compromise text):**
> **Article 14a(2)(h)** states: *"the existence of automated decision making including profiling referred to in Article 20(1) and (3) and at least in those cases, meaningful information about the logic involved, as well as the significance and the envisaged consequences of such processing for the data subject."* [37]

However, where black box algorithms are concerned – are data controllers able to fully explain the logic involved in their automated decision-making? While there are technical methods available (e.g. [38]), they mostly say: This feature (e.g. income) weighs $x$, and as a result you have been assigned to tariff-A as opposed to tariff-B. This is not a logically crisp statement. Furthermore, can data subjects be certain that such logic explanations will be expressed in terms that the layperson understands?

Moreover, similar to the protected characteristics outlined by the Equality Act, Article 9 of the GDPR outlines several "special categories of personal data":

> **GENERAL DATA PROTECTION REGULATION (2015 compromise text):**
> *"Article 9 Processing of special categories of personal data*
> *1. The processing of personal data, revealing racial or ethnic origin, political opinions, religious or philosophical beliefs, trade-union membership, and the processing of genetic data, biometric data in order to uniquely identify a person or data concerning health or sex life and sexual orientation shall be prohibited. […]"* [37]

Article 9 therefore includes some categories of data that are not listed as protected characteristics by section 4 of the Equality Act, e.g. biometric and genetic data. However, it does not explicitly mention some of these protected characteristics, e.g. age, gender reassignment, and marriage and civil partnership. Article 9(2) also provides a ten exceptions ((a)-(i)) to this prohibition, including where a data subject explicitly consents to the processing of such data (Article 9(2)(a)) and for reasons of public interest in the area of public health (Article 9(2)(hb)).

---

[16] For instance, David Wright et al. [102, p. 243] state: "Non-discrimination rules should be taken into consideration at the design stage of technology and service development."

**Some other key points about profiling and the GDPR:**

- In general, there may not be many major differences between Article 15 of the Data Protection Directive and Article 20 of the GDPR [39];
- Article 20 (GDPR) expands the list of profiling examples found in Article 15 (DPD) [39];
- The "significantly affects" test present in Article 20 seems to set a very high threshold (e.g. would a marketing decision be said to significantly affect an individual?) [39] – however, unlawful discrimination is highly likely to fall under this category;
- Businesses need explicit consent rather than unambiguous consent for automated profiling in many instances [39];
- Article 14(1) of the GDPR – data subjects not a right to avoid profiling but avoid being subject to a completely automated decision [40];
- Article 20(3) of the GDPR refers to "special categories of personal data" [40] which are comparable to the "protected characteristics" under the Equality Act; and,
- Under Article 19 of the GDPR, data subjects have the right to object even where the profiling is lawful [40].

### 2.3.2 Accurate data

There is also uncertainty over the extent in which automated decision-making systems have to utilise accurate data. For instance, in the case of *Smeaton v Equifax Plc* [2013] EWCA Civ 108, the UK's Court of Appeal held that credit reference agencies are not obliged to ensure absolute data accuracy under the UK Data Protection Act 1998; although they should take reasonable steps to ensure that it is up-to-date. In this case, Mr Smeaton had a bankruptcy order rescinded in May 2002. However, his later application for a business account and loan at a bank was rejected, as the bankruptcy order had not been removed from his credit file by Equifax. Equifax obtained its bankruptcy data from the London Gazette (where bankruptcies must be advertised). However, the London Gazette does not necessarily contain advertisements of annulments or rescindments; therefore Equifax did not up-date its data as no rescindment was published in the Gazette. Furthermore, it was held that it was neither reasonable for Equifax to have identified this "blind spot" or lobbied the Government to change the legislative framework.

However, note that in the landmark "right to be forgotten" ruling (– taken by the European Court of Justice in the case of *Google Spain SL v Agencia Espanola de Proteccion de Datos (AEPD)* (C-131/12) [2014] 3 W.L.R. 659 which centred on auction notice of a repossessed home –) did give rise to a duty to process accurate and timely data.

In the USA, a FTC study [41] also found that credit scores are not without error:

> *"[…] five percent of consumers had errors on one of their three major credit reports that could lead to them paying more for products such as auto loans and insurance. Overall, the congressionally mandated study on credit report accuracy found that one in five consumers had an error on at least one of their three credit reports."*

In consequence, New City Council were considering a bill that would prevent employers using credit histories for purposes of hiring new employees [42]. Therefore, should more be done to protect data subjects from inaccuracies? Will the GDPR be able to strengthen individuals' rights through the right to rectification under Article 16?

> **GENERAL DATA PROTECTION REGULATION (2015 compromise text)**
> **Article 16 of the GDPR** (2015 compromise text version) states: *"The data subject shall have the right to obtain from the controller without undue delay the rectification of personal data concerning him or her which are inaccurate. Having regard to the purposes for which data were processed, the data subject shall have the right to obtain completion of incomplete personal data, including by means of providing a supplementary statement."* [37]

### 2.3.1 Example v – credit referencing scores

In the UK, there the three main companies – CallCredit, Experian and Equifax – that compile information on how consumers manage their credit and payments [43]. This information is one method used by banks and lenders to determine the level of risk a particular consumer poses (i.e. low risk = good credit score, high-risk = bad credit score) [43].[17] For instance, missed child maintenance payments may damage a consumer's credit rating [44]. This credit score includes information such as: number of active bank accounts, court judgments, people who are financially linked to the consumer in question (e.g. a joint bank account), current and previous addresses, and whether the consumer entered on the electoral register [43]. The sources of data used may be expanding, for instance the use of social media information [45].

The Guide to Credit Scoring 2000 [46] was established by a multitude of organisations in the UK credit industry.[18] It provides a guide to "all the major developers and users of credit scoring systems" [46, p. 3]. Point 2.4 of the Guide [46, p. 6] clearly states: *"Credit Scoring will not discriminate on the grounds of sex, race, religion, disability or colour. All scoring systems will be designed and used in a way that conforms to all relevant legislation."*

However, in the USA, there have been allegations that credit scoring may involve indirect discrimination i.e. credit scores are used a proxies for a number of protected characteristics such as race. For instance, Sarah Ludwig [47] asserts that "credit scores and reports are not race neutral", and ethnic minorities are "disproportionately targeted for high-cost, predatory loans".

A US report conducted by Robert B. Avery et al. [48, p. 3] in 2010 focused on better-understanding whether credit scoring causes a disparate impact to people with specific protected characteristics i.e. race or ethnicity, age, and/or gender. Avery et al. [48, p. 7] used a nationally representative sample of that comprised over 300,000 anonymous credit records observed during June 2003 and December 2004; these data were supplemented by demographic information for each individual from the Social Security Administration and a demographic information company. The results of this study suggest: "that credit scores do not have a disparate impact across race, ethnicity, or gender" [48, p. 26]. However, it also found: "some evidence that credit characteristics associated with the length of an individual's credit history […] may have a disparate impact by age" [48, p. 26].

---

[17] The Guide to Credit Scoring 2000 [46, p. 11] defines credit score: *"The sum of the points calculated within the credit scorecard gives the credit score"* and credit scorecard: *"A credit scorecard consists of a set of attributes and the points (or weightings) assigned to them."* For more background information on credit scores see: [120].

[18] Association for Payment Clearing Services, British Bankers Association, Building Societies Association, Consumer Credit Trade Association, Council of Mortgage Lenders, Credit Card Research Group, Finance & Leasing Association, Institute of Credit Management and Mail Order Traders Association; together with: Consumer Credit Association (UK), Equifax, Experian, and Scorex (UK) Ltd.

> Given the public interest in this area, there appears to have been limited research, Avery et al. [48, p. 2] provide two main reasons for this: (1) credit scoring is largely proprietary i.e. there is very limited/no access to detailed information about how these models work in practice; and (2) credit scoring does not use protected attributes (e.g. race and gender) and therefore no explicit links between credit scores, credit history and demographic information.[19] In consequence, the extent in which automated discrimination is prevalent within the insurance industry appears unclear.
>
> While particular attention has been given to credit scoring and discrimination from the field of economics, it has received less consideration from law – a point highlighted by Federico Ferretti [49]. Ferretti [50, p. 268] states:
>
> *"In fact, it seems that the expanding use of credit reporting, coupled with the many different activities that CRAs carry on, does not find limits in today's law, thus being open to further new technological advances and findings. Ultimately, it would be worth exploring whether specific regulation of this market is necessary to avoid the arbitrary use of credit reference information and the arbitrary positioning of privately owned CRAs in the modern society."*[20]

### 2.3.2 Accountability and transparency

With these limitations in mind – what is the best way forward for the legal framework? A key issue is making these automated decision-making systems more accountable to the people they are profiling. Greater transparency could be achieved by placing a limited duty of disclosure on those responsible for such automated decision making. This may involve the release of (redacted) input and output data, and a discrimination impact assessment (perhaps an addition to the data-protection-by-design and by default obligations and data protection impact assessments set out by Articles 23 and 33 of the European General Data Protection Regulation). However, at the same time, it must be considered whether it is possible to formulate such a duty without jeopardising the owners' intellectual property rights.[21]

### 2.4 Summary: relationship between equality and data protection

Bart Custers [51] makes an important distinction between the protections granted to individual profiling under data protection law, in contrast to group profiling. Custers [51, p. 291] defines individual and group profiling as follows:

> *"A personal profile is a property or a collection of properties of a particular individual. Profiles concerning a group of persons are referred to as group profiles. Thus, a group profile is a property or a collection of properties of a particular group of people."*

Given that the very nature of individual profiling involves personally identifiable data and therefore falls under the scope of data protection law, is there the potential for group profiling to fall outside is scope where personal data are sufficiently anonymised [51, p. 292]? However, what if this group profiling information is then used to unfairly discriminate against certain groups? Is this where equality law is needed to step in?

---

[19] Furthermore, there have been studies that show minor discrepancies between men and women's credit scores [112].

[20] For more information about the regulation of credit rating agencies see [108] and [109].

[21] Mireille Hildebrandt and Bert-Jaap Koops [123, pp. 440-441] state: *"As profiling processes are frequently covered by trade secret provisions or intellectual property rights, and because the technology may be a black box even for the data processor, adequate supervision of the duty to register is unachievable. […] More often than not, non-compliance with data protection law does not occur deliberately but is caused by a lack of knowledge or understanding of the legislation, since the rules are unknown, ambiguous, vague, or too complex to be comprehensible."*

It appears that further clarification is required on the exact relationship between equality law and data protection law. Daniel Le Métayer and Julien Le Clainche [52, p. 327] "advocate the establishment of stronger connections between anti-discrimination and data protection laws, in particular to ensure that any data processing resulting in unfair differences of treatments of individuals is prohibited and is subject to effective compensations and sanctions." Eric E. Schadt [53, p. 612] further states: "To adapt to this rapidly changing social and technological landscape in ways that serve our individual best interests and that of society more generally, I believe education and legislation aimed less at protecting privacy and more at preventing discrimination will be key."

As the digital age matures and automated decision-making becomes further entrenched in our daily lives, these two areas of law may need to become more explicitly connected in order to confront automated unlawful discrimination; this therefore constitutes an important area of interest for future work.

## 3. THE TECHNOLOGICAL APPROACH TO AUTOMATED DISCRIMINATION

### 3.1 Introduction

In an attempt to infer whether there has been any (unlawful) discrimination, the starting point for most discrimination aware approaches is to use statistics. In the field of economics, this discrimination discovery methodology is known as "statistical discrimination" [54, p. 260].[22] However, a simple statistical check is not always sufficient; for instance, it may overlook instances of individual discrimination,[23] positive action, and Simpson's Paradox (see below). Furthermore, some of those involved with analysing (big) data may commit statistical errors[24] (e.g. oversimplification of results and irrelevant sampling) if they do not possess/apply a sufficient level of statistical proficiency [16].

Therefore, in order to more effectively safeguard data subjects from unlawful discrimination, there is a pragmatic need for: (a) application of a sufficient level of statistical proficiency by those analysing data; and (2) further robust checking procedures that go beyond mere statistical analysis – one such area is discrimination-aware data-mining (DADM).

### 3.2 Example vi – Simpson's Paradox: Admissions to University of California, Berkeley

P.J. Bickel et al. [55] sought to analyse whether the decision to select applicants to the University of California, Berkeley in autumn 1973 was influenced by gender. The University received 12,763 complete applications – out of these applications: 8442 were submitted by male applicants; and 4321 were submitted by female applicants [55, p. 398]. Bickel et al. started with the "simplest approach" – to aggregate the data for the entire campus. In total around 44% of the male applicants were admitted to the University, whilst around 35% of the females were admitted [55, p. 398]. At first glance, it appeared there had been some form of discrimination based on gender – however, this was later found to be misleading. This was because the initial statistical analysis did not take into account the underlying dependencies, i.e. how the entry process to different departments varied. For instance, some departments attract more or less female applicants than male applicants, e.g. only 2% of applicants to mechanical engineering were female [55, p. 399].

In consequence, the second approach of Bickel et al. was to use disaggregation, i.e. examine the data of each of the departments individually [55, p. 400]. In fact, the majority of departments showed a "small but statistically significant bias" in favour of female applicants [55, p. 403]. This is therefore an example of Simpson's Paradox where a trend that appears in a number of groups (e.g. the departments) disappears or reverses when these data are combined (e.g. the overall applications for the University).

---

[22] For an example of the use of statistics within discrimination discovery see [115].

[23] Andrea Romei et al. [93, p. 6064] state: "*In the socio-economic field the problem has been addressed by analysing data with a statistical approach. The basic idea is to see, by means of regression analysis, whether sensitive features, like gender and race, are correlated with a less favorable treatment of individuals. […] As an example, consider the case of loan applications to a bank. The discriminatory behavior of a single branch office against applicants from a local minority can readily be hidden in the much larger set of decisions of the whole bank. In a few words, the statistical approach tends to find a general model characterizing the whole population, whereas discrimination often arises in specific contexts.*"

[24] For instance, Google Flu Trends [134].

(Note that Indrė Žliobaitė et al. [56, pp. 994-995], Salvatore Ruggieri [57, p. 878] and Francesco Bonchi et al. [58, p. 2] all raise the importance of investigating discrimination outcomes and cite this example.)

### 3.3 Discrimination-aware data-mining (DADM)

The emerging area of discrimination-aware data-mining (DADM) [25] (also known as discrimination-aware learning [59, p. 1532], fairness-aware, and discrimination discovery in databases) is focused on "finding unfair practices against minorities which are hidden in a dataset of historical decisions" [60, p. 1127]. DADM is **"discovery-driven"** and therefore differs from the **"assumption-driven"** approach of statistical discrimination, as Roderick M. Rejesus [61, p. 24] further explains:

> "Data mining differs from traditional statistics in the following way — formal statistical inference is 'assumption-driven', in the sense that a hypothesis is first formed and then validated against data. On the other hand, data mining is 'discovery-driven', in the sense that patterns and hypothesis are automatically extracted from the data."

Deepali Jagtap and Shirish S. Sane outline the three ways in which discrimination (as a result of data mining) can be prevented [62, p. 29]: (1) **the pre-processing method** i.e. where "the original dataset is modified so that it will not result in discriminatory classification rule" [62, p. 29]; (2) **the in-processing method** i.e. "where data mining algorithm is modified instead of modifying original dataset" [62, p. 30]; and (3) **the post-processing method** i.e. where the "resultant mining model is modified instead of modifying original data or mining algorithm" [62, p. 30].

### 3.3.1 Table of discrimination discovery approaches

The following table provides a brief overview of some of the key literature in the area of discrimination-aware data-mining (DADM). Please note this is by no means an exhaustive list; refer to original articles for full information.

| Year | Author(s) | Brief description of approach | Example(s)/test data used |
| --- | --- | --- | --- |
| 2008 | *Dino Pedreshi, Salvatore Ruggieri and Franco Turini* [63][26] | Introduces the notion of discriminatory classification rules, which are split into two groups: (1) potentially discriminatory rules (PD) – e.g. gender and ethnic minority; and, (2) potentially non-discriminatory rules – e.g. rarely offer credit to people living in a certain postcode. | German credit case study |
| 2010 | Kamiran, F.; Calders, T.; Pechenizkiy, M.; [64][27] | "[P]ropose the construction of decision trees with non-discriminatory constraints" [64, p. 874]. | Income data; Dutch Census datasets 1971 and 2001 |
| 2010 | *Salvatore Ruggieri,* | "We present a reference model for finding (prima facie) | German credit case study; |

---

[25] F. Kamiran et al. [64, p. 869] define discrimination-aware: "The solution is to develop new techniques which we call discrimination aware — we want to learn a classification model from the potentially biased historical data such that it generates accurate predictions for future decision making, yet does not discriminate with respect to a given discriminatory attribute." Andrea Romei et al. [93, p. 6066] define discrimination discovery in databases as: "[…] the actual discovery of discriminatory situations and practices hidden in a large amount of historical decision records. The aim is to unveil contexts of possible discrimination on the basis of legally-grounded measures of the degree of discrimination suffered by protected-by-law groups in such contexts."

[26] This appears to be one of the first articles published in the area of discrimination-aware data mining [63, p. 560].

[27] Available at <doi.org/10.1109/ICDM.2010.50>

| | | | |
|---|---|---|---|
| | *Dino Pedreshi and Franco Turini [65]*[28] | evidence of discrimination in automatic DSS which is driven by a few key legal concepts. First, frequent classification rules are extracted from the set of decisions taken by the DSS over an input pool dataset. Key legal concepts are then used to drive the analysis of the set of classification rules, with the aim of discovering patterns of discrimination. We present an implementation, called LP2DD, of the overall reference model integrating induction, through data mining classification rule extraction, and deduction, through a computational logic implementation of the analytical tools." [65, p. 157] | Mark<br>Musa Hussein v Saints Complete House Furnishers 35257/78 1979 WL 466761 |
| 2010 | *Salvatore Ruggieri, Dino Pedreshi and Franco Turini [60]*[29] | A demonstration of the DCUBE system for discrimination discovery, which could be used "*by a large audience of users, including owners of socially-sensitive decision data, government antidiscrimination analysts, technical consultants in legal cases, researchers in social sciences, economics and law.*" [60, p. 1127] | Two publically available loan datasets:<br>German credit dataset;<br>PKDD 1999 financial dataset |
| 2010 | *Salvatore Ruggieri, Dino Pedreshi and Franco Turini [66]* | Focuses on classification rules | German credit dataset and on the PKDD Discovery Challenge 1999 financial dataset |
| 2010 | *Toon Calders and Sicco Verwer [67]* | Focuses on three naive Bayes approaches for discrimination-free classification | |
| 2011 | *Indrė Žliobaitė, Faisal Kamiran and Toon Calders [56]* | Focuses on the design of discrimination-free classifiers in the context of direct and indirect discrimination | Admission to fictitious University;<br>2001 Dutch Consensus; |
| 2011 | *Toshihiro Kamishima, Shotaro Akaho and Jun Sakuma [1]*[30] | "*In this paper, we first discuss three causes of unfairness in machine learning. We then propose a regularization approach that is applicable to any prediction algorithm with probabilistic discriminative models. We further apply this approach to logistic regression and empirically show its effectiveness and efficiency.*" [1, p. 643] | Adult/Census Income distributed at the UCI Repository |
| 2011 | *Binh Thanh Luong, Salvatore Ruggieri and Franco Turini [68]*[31] | "With the support of the legally-grounded methodology of situation testing, we tackle the problems of discrimination discovery and prevention from a dataset of historical decisions by adopting a variant of k-NN classification." [68, p. 502] | German credit dataset |
| 2012 | *Cynthia Dwork, Moritz Hardt, Toniann Pitassi, Omer Reingold and Richard Zemel [69]*[32] | "We study fairness in classification, where individuals are classified, e.g., admitted to a university, and the goal is to prevent discrimination against individuals based on their membership in some group, while maintaining utility for the classifier (the university). The main conceptual contribution of this paper is a framework for fair classification comprising (1) a (hypothetical) task-specific metric for determining the degree to which individuals are similar with respect to the classification task at hand; (2) an algorithm for maximizing utility subject to the fairness constraint, that similar individuals are treated similarly." [69, p. 214] | University admission |
| 2012 | *Dino Pedreshi, Salvatore Ruggieri and Franco Turini [70]*[33] | Focuses on top-k measures for discrimination discovery | German credit dataset |
| 2012 | *Toshihiro Kamishima, Shotaro Akaho, Hideki Asoh and Jun Sakuma [71]*[34] | "In this paper, we first discuss three causes of unfairness in machine learning. We then propose a regularization approach that is applicable to any prediction algorithm with probabilistic discriminative models. We further apply this approach to logistic regression and empirically show its | Adult / Census Income distributed at the UCI Repository |

---

[28] Available at <http://delivery.acm.org/10.1145/1570000/1568252/p157-pedreschi.pdf?ip=152.78.38.231&id=1568252&acc=ACTIVE%20SERVICE&key=BF07A2EE685417C5%2EA13CBF7F1C3C7DF4%2E4D4702B0C3E38B35%2E4D4702B0C3E38B35&CFID=582465740&CFTOKEN=91902579&__acm__=1455280926_e334b4dc11c7bee0a3e59862f40eb5ef> [last accessed 12 February 2016].

[29] Available at <http://dl.acm.org/citation.cfm?id=1807298&CFID=582465740&CFTOKEN=91902579> [last accessed 12 February 2016].

[30] Available at <http://ieeexplore.ieee.org/stamp/stamp.jsp?tp=&arnumber=6137441> [last accessed 12 February 2016].

[31] Available at <http://dl.acm.org/citation.cfm?id=2020488> [last accessed 12 February 2016].

[32] Available at <http://dl.acm.org/citation.cfm?id=2090255> [last accessed 12 February 2016].

[33] Available at <dl.acm.org/ft_gateway.cfm?id=2245303> [last accessed 12 February 2016].

[34] Available at <http://link.springer.com/chapter/10.1007%2F978-3-642-33486-3_3> [last accessed 12 February 2016].

| | | effectiveness and efficiency." [71, p. 35] | |
|---|---|---|---|
| 2013 | Goce Ristanoski, Wei Liu and James Bailey [59][35] | "Our work focuses on an aspect often overlooked in the discrimination aware classification - the scenario of an imbalanced dataset, where the number of samples from one class is disproportionate to the other. We also investigate a strategy that is directly minimizing discrimination and is independent of the class balance." [59, p. 1529] | Adult and census income; German credit data |
| 2013 | Salvatore Ruggieri [57][36] | "We investigate the relation between t-closeness, a well-known model of data anonymization, and α-protection, a model of data discrimination." [57, p. 875] | University admission; German credit data |
| 2014 | Bettina Berendt and Sören Preibusch [72] | Conduct a large-scale experimental user study to test the utility of discrimination-aware data-mining tools. | Bank loan scenario |
| 2014 | Deepali Jagtap and Shirish S. Sane [62][37] | Focuses on discrimination discovery and data transformation. | German credit data |
| 2014 | Koray Mancuhan and Chris Clifton [73][38] | Proposes a discrimination discovery and a non-discriminatory classification technique using Bayesian networks, which aims to capture instances of direct and indirect discrimination | German Credit Dataset; US Census Income Dataset from the UCI repository |
| 2015 | Francesco Bonchi, Sara Hajia, Bud Mishra and Daniele Ramazzotti [58] | Focuses on a principled causal approach to discrimination discovery through algorithms | German credit data; Census income data; Berkeley Admissions data |
| 2015 | Michael Feldman, Sorelle A. Friedler, John Moeller, Carlos Scheidegger and Suresh Venkatasubramanian [54][39] | "We present four contributions. First, we link disparate impact to a measure of classification accuracy that while known, has received relatively little attention. Second, we propose a test for disparate impact based on how well the protected class can be predicted from the other attributes. Third, we describe methods by which data might be made unbiased. Finally, we present empirical evidence supporting the effectiveness of our test for disparate impact and our approach for both masking bias and preserving relevant information in the data." [54, p. 259] | German credit data; Adult income datasets |
| 2015 | Haijan [3] et al. | Investigates: "the problem of discrimination and privacy-aware frequent pattern discovery, i.e., the sanitization of the collection of patterns mined from a transaction database in such a way that neither privacy-violating nor discriminatory inferences can be inferred on the released patterns. […] We also proposed an algorithm to take into account the legal concept of genuine requirement to make an original pattern set protected only against unexplainable discrimination." [3, p. 1779] | German credit dataset |

### 3.4 Technological approach summary

While Murielle Hildebrandt [74, p. 193] makes clear that such DADM technologies are yet to have a major impact, they do appear to have significant future potential as a tool for confronting (in)direct automated discrimination. Furthermore, the merits of an interdisciplinary approach to discrimination prevention and discovery in the context of data analytics has already been raised by Salvatore Ruggieri et al. [66, p. 38]:

> "Clearly, many issues in discrimination-aware data mining remain open for future investigation, both on the technical and on the interdisciplinary side. […] On the

---

[35] Available at <http://dl.acm.org/citation.cfm?id=2507836> [last accessed 12 February 2016].
[36] Available at <http://ieeexplore.ieee.org/xpls/abs_all.jsp?arnumber=6754013> [last accessed 12 February 2016].
[37] Available at <http://research.ijcaonline.org/volume95/number25/pxc3897037.pdf> [last accessed 11 February 2016].
[38] Available at <http://link.springer.com/article/10.1007%2Fs10506-014-9156-4> [last accessed 10 March 2016].
[39] Available at <http://dl.acm.org/citation.cfm?id=2783311&CFID=751810830&CFTOKEN=74691654> [last accessed 12 February 2016].

> *interdisciplinary side, it is important to pursue the interplay with legislation and regulatory authorities."*

In addition, as is the case with the use of statistical analysis, concerns have been raised by Francesco Bonchi et al. [58, p. 11] that some of the DADM research focuses on correlation rather than causation:[40]

> *"Discrimination discovery from databases is a fundamental task in understanding past and current trends of discrimination, in judicial dispute resolution in legal trials, in the validation of micro-data before they are publicly released. While discrimination is a causal phenomenon, and any discrimination claim requires to prove a causal relationship, the bulk of the literature on data mining methods for discrimination detection is based on correlation reasoning."*

Therefore, it is important that automated discrimination approaches are aware that "correlation does not imply causation", and include some mechanism for sufficient statistical proficiency.

---

[40] Kiran Soar [113] offers a legal perspective on "A probabilistic theory of legal causation".

## 4. PROPOSING A NEW INTERDISCIPLINARY APPROACH TO AUTOMATED DISCRIMINATION

### 4.1 Introduction

This feasibility study has begun to examine some of the key interdisciplinary literature concerned with the legal and technological approaches to automated discrimination (see Sections 2 and 3). The research shows that a reliance on a quantitative approach alone (e.g. a simple statistical check) is not enough to confront automated discrimination, as it may overlook instances of Simpson's Paradox, positive action and individual discrimination. For that reason a mixed approach of quantitative and qualitative elements is necessary; e.g. examination of automated decision-making on an individual basis, ethics procedures and stakeholder analysis. Furthermore, given that not all instances of discrimination will be unlawful, there is need to strengthen the technological approach of DADM with legal compliance checking. In consequence, the research has shown that there is a pragmatic need for an interdisciplinary approach that brings together these legal and technological elements [66, p. 38].

This section therefore aims to build on this research by making preparations for further development of an interdisciplinary methodology, which should offer those directly involved with the design, development and utilisation of black box algorithms, a robust checking procedure that: (1) effectively safeguards data subjects from unlawful discrimination; and (2) helps those data controllers utilising automated decision-making to meet their legal obligations.

This discrimination-aware approach needs to begin from the very outset of the design and development of the black-box data crunching algorithm, i.e. a discrimination-aware culture is required. In order to provide both proactive and reactive discrimination checks, the inputs, outputs and internal workings of the black box algorithm (if and to what extent they can be revealed) must be investigated. Key ethical questions need to be addressed, such as is the purpose of this automated decision-making justified?  Has best practice been implemented (e.g. industry standards, in-house policy and ethical guidelines)? This methodology therefore should aid: (1) **discrimination-prevention**, i.e. be proactive; and (2) **discrimination-discovery** i.e. be reactive.

An essential part of this methodology will be focused on developing a model that represents norms describing undesirable biases against sensitive attributes and metrics for describing situations of norm violation and compliance. In other words, a model that formally describes whether a certain result (e.g. the number of female employees hired in comparison to male employees) constitutes an unlawful discrimination.

This section therefore puts forward a plan for such an interdisciplinary methodology to be taken forward in a future follow-up project.

### 4.2 Key points to consider

i. **The inputs to the black box algorithm must be scrutinised**
Those directly involved with the design, development and employment of such automated decision-making systems need to consider and minimise the potential impact of any inaccuracies, past/current prejudices and/or bias that could potentially result in (in)direct discrimination. For instance, what is the quality and provenance of the data used (this is especially important when a number of datasets are from third party sources)? Are these data fit for purpose – e.g. are they timely and have they been gathered by an impartial source? Is there a high risk that historical prejudice is present in (some of) the datasets?

ii. **The outputs to the black box algorithm must be examined**
Although a starting point for assessing whether unlawful discrimination has taken place, in the majority of cases a simple statistical check is not sufficient. For instance, overall statistics may overlook instances of positive action or a case of Simpson's Paradox.

iii. **Statistical robustness**
Those directly involved with the design, development and employment of such automated decision-making systems need to ensure that best practice is followed when utilising statistical methods. This is in order to minimise statistical error and prevent oversimplification (i.e. "correlation does not imply causation").

iv. **The selected attributes need to be assessed**
Those directly involved with the design, development and employment of such automated decision-making systems need to consider what potential (and obvious) loopholes could be used to serve as proxies for protected characteristics, e.g. the category "job title" could be used as a proxy for gender. For instance, have any protected characteristics been used? Is it likely that, in the absence of protected characteristics, proxies have been used? Does any of the input data provide a source of background data (e.g. demographics information)?

v. **Due diligence**
Due diligence is also an essential requirement – there is need to keep up-to-date with legal developments. Equality monitoring data could be used as a potential tool to further assist with discrimination assessment.

vi. **The representation of fairness and unfairness as explicit norms**
Socio-cultural, legal and other norms have been explicitly represented in agent-based systems. In some cases, they have been represented as specific logics (e.g. teleological [41] and deontic [42]). However, by restricting such representations to propositional logics, there is no means of describing discrimination. This therefore leads onto the first crucial question: in what ways is it possible to formally describe that a certain result (e.g. the number of male and female applicants hired by the three companies in Scenario A below) is either discriminatory or non-discriminatory? The second crucial question is: on the basis that a certain result is formally described as discriminatory, is this discrimination unlawful?

vii. **Internal training, policies and procedures**

---

[41] Oxford English Dictionaries Online [127] define "teleology": "*Philosophy* The explanation of phenomena by the purpose they serve rather than by postulated causes."

[42] Oxford English Dictionaries Online define "deontic": "*Philosophy* Relating to duty and obligation as ethical concepts."

Are staff members provided with robust training on anti-discrimination practices? Is there any evidence of a data protection impact assessment/data-protection-by-design from the outset of the design and development of the black box algorithm? Should the process be fully automated? Should there be someone responsible for overseeing the process on a daily basis?

viii. **Automated discrimination assistance tools**
What tools are available to those directly involved with the design, development and employment of such algorithms to automatically source the potential areas of discrimination? What tools need to be developed? How can DADM be better taken forward in a practical sense?

ix. **Third party inspection**
These checking procedures must be open to some form of third party scrutiny to ensure their effectiveness, e.g. (as earlier mentioned) by placing a limited duty of disclosure on those responsible for such automated decision-making.

### 4.3 Scenarios to consider

It is anticipated that this future methodology would be able to assist with answering the types of questions raised in the two following scenarios:

---

**SCENARIO A – JOB APPLICANTS:**
Three companies advertise 100 vacant employment positions. Each company receives 500 applications from male applicants and 500 applications from female applicants.
*Company 1* hires: 51 men and 49 women.
*Company 2* hires: 70 men and 30 women.
*Company 3* hires: 90 men and 10 women.
**The key question is: which of these three companies, if any, has unlawfully discriminated based on gender?**

---

**SCENARIO B – *ACME INSURANCE CO.*:**

Despite no significant changes to the European health insurance market over the past year, *ACME Insurance* Co. (a fictional health insurance provider) notices a significant decrease in new customers signing up for its health insurance policies during January to February 2016. All *ACME Health Insurance* Co.'s insurance tariffs are very competitively priced, appear on popular comparison websites, and are widely-marketed through leading media channels. For the past 10 years, *ACME Health Insurance* Co. has been a respected market leader in European health insurance. *ACME Health Insurance* Co. dispatches a team of employees to investigate this change.

The investigative team focus on *ACME Health Insurance* Co's automated online quote system. While the amount of people applying for a quote as significantly increased over the past two months, less people have signed up to its tariffs. The investigative team notice that there has been a disproportionate rise in the amount of customers being matched to its most expensive tariffs.

*ACME Health Insurance* Co notices that a small number of individuals are starting to complain about the health insurance quote they obtained from the company. The majority of those criticising the company are women, and received quotes for the highest tariffs. A number of men have responded to these women stating that they were offered a reasonable quote from *ACME Health Insurance* Co. The investigative team also find that during January-February 2016 only 17% of new customers were woman in contrast to the monthly average 52% (calculated over the past ten years). *ACME Health Insurance* Co. clearly states in its terms and conditions that: "we do not use gender to determine your tariff".

*ACME Health Insurance* Co. is now deeply concerned that its automated online quote system is unconsciously discriminating against its prospective female customers. *ACME Health Insurance* Co. immediately suspends its automated online quote system until further notice.

**The key questions are: (1) how can *ACME Health Insurance* Co. quickly determine and bring an end to the source(s) of this unconscious discrimination? (2) What potential liabilities does ACME Health Insurance Co. now face? (3) How can *ACME Health Insurance* Co. make its automated online quote system more robust in the future to prevent this from happening again?**

## 5. PROJECT SUMMARY AND FUTURE WORK

### 5.1. Conclusions

The research question this project aims to confront is: how, and to what extent, those directly involved with the design, development and employment of a specific "black box" algorithm can be certain that it is not unlawfully discriminating (directly and/or indirectly) against particular persons with protected characteristics (e.g. gender, race and ethnicity)? This feasibility study has shown that it is of paramount importance that this particular question is addressed. However the provision of such anti-discrimination checks and guarantees is fraught with difficulties where black box algorithms are involved (see Section 1). While software engineers have access to inputs (e.g. third party data) and outputs (e.g. patterns and trends) of a specific algorithm, they often are not privy to its inner workings (i.e. exactly how the algorithm derives outputs from the inputs). Furthermore, the current legal framework provides a limited approach with regard to automated discrimination (see Section 2).

While the interdisciplinary methodology outlined in this article is no panacea, it could potentially help those directly involved with the design, development and utilisation of black box algorithms to better-safeguard data subjects from discrimination. Furthermore, greater accountability and transparency could better inform data subjects about potential instances of unlawful discrimination.

### 5.2 Recommendations for future work

#### 5.2.1 The involvement of wider interdisciplinary expertise

Automated discrimination is a complex and multi-faceted issue that cannot be addressed by one discipline in isolation. In consequence, a future research project would benefit from a wider interdisciplinary team, including legal, technological, social-cultural, statistical and interdisciplinary (e.g. web science) experts.

|   | Discipline (a-z) | Some (potential) key contributions |
|---|---|---|
| 1 | **Computer science** (e.g. data miners and black box testers) | <ul><li>To provide in-depth understanding of data mining, DADM, black box testing, other relevant discrimination discovery and prevention tools and technical industry standards;</li><li>The ability to design, build and test computation tools to discover and prevent automated discrimination; and,</li><li>To help to reach a common interdisciplinary understanding of discrimination by providing the technological context.</li></ul> |
| 2 | **Humanities** (e.g. historians and philosophers) | <ul><li>To offer a better-understanding about how the notion of discrimination has developed – and to what extent current issues are (not) a re-manifestation of past issues;</li><li>To examine what approaches to discrimination prevention and detection have been used before – what aspects are still important in any future methodology?</li><li>To help to reach a common interdisciplinary understanding of discrimination by providing the socio-cultural context.</li></ul> |
| 3 | **Law** (e.g. equality, ICT, intellectual property and insurance law) | <ul><li>To provide in-depth understanding of the legal framework concerning automated discrimination from related legal areas and multi-jurisdictions;</li><li>To offer guidance on legal compliance; and,</li><li>To help to reach a common interdisciplinary understanding of discrimination by providing the legal context.</li></ul> |
| 4 | **Relevant interdisciplinary researchers** | <ul><li>To provide a range of disciplinary skills (from the other required areas); and,</li></ul> |
|   | (e.g. web science) | ▪ To approach automated discrimination from a number of perspectives. |
| 4 | **Social sciences** (e.g. sociology, psychology and economics) | ▪ To offer qualitative and quantitative expertise e.g. stakeholder analysis, (large-scale) experiments, interviewing and survey techniques. This could be used to assist with a better-understanding of the attitudes and perceptions of data subjects and those directly involved with the design, development and utilisation of black box algorithms over automated discrimination;<br>▪ To help to reach a common interdisciplinary understanding of discrimination by providing the socio-cultural context; and,<br>▪ For enhanced knowledge of statistical approaches to discrimination. |
| 5 | **Statistics** (E.g. mathematics and economics) | ▪ To examine: (1) the extent in which statistical analysis should be used for automated discrimination prevention and discovery; (2) the most beneficial types of statistical analysis in this context; and, (3) best practice – including any required safeguards; and,<br>▪ To help to reach a common interdisciplinary understanding of discrimination by providing the analytical context. |

### 5.2.2 The enhancement of disciplinary approaches to automated discrimination (sections 2-3)

The following table provides a list of five key recommendations that require further research:

|   | **Recommendations** | **Brief overview** |
|---|---|---|
| 1 | **Improved definition of automated discrimination**<br><br>[**Law + Socio-cultural**]* | This report has focused on how "unlawful" discrimination is defined within UK law. However, discrimination is not a fixed concept. For instance, music recommendation systems may be welcomed, but other types of discrimination can be undesirable and even abhorrent. Does the current legal definition go far enough to cover all major instances of automated discrimination? Is a new interdisciplinary definition of automated discrimination required?<br><br>The perception of discrimination as lawful and unlawful depends on societal attitudes of a particular time and place. Therefore, discrimination is not a fixed concept, but its parameters change with developing societal attitudes. Furthermore, what constitutes lawful discrimination in one jurisdiction may be unlawful in another (e.g. offering female drivers more preferential tariffs for car insurance than men).<br><br>Different disciplines often differ on the meaning of discrimination. Therefore, it is of importance that socio-cultural experts provide a greater understanding of the development of the term discrimination, and help the project move towards a common understanding of lawful and unlawful discrimination. Furthermore, it is important that the methodology is able to distinguish between legal, technical, historic, socio-cultural and jurisdictional definitions of discrimination. |
| 2 | **Greater mapping between equality law, data protection law and other relevant areas**<br><br>[**Law**]* | Section 2 of this report outlines some key legal issues to consider in regard to automated discrimination. One key area for future research raised was gaining a better-understanding of how equality law and data protection law overlap in practice. Is there a need for both areas to become more compatible? |
| 3 | **Focus on multi-jurisdictional approaches to automated discrimination**<br><br>[**Law + Socio-cultural**]* | The cross-jurisdictional nature of automated discrimination must be taken into account. For instance, what happens when a person situated in one country is automatically discriminated against by a system operating and based in another country? Furthermore, how does the socio-cultural context in which potential automated discrimination is taking place affect the attitudes of people across jurisdictions? Finally, what lessons can be learnt from other jurisdictional approaches to the prevention and discovery of automated discrimination? |
| 4 | **Focus on black box testing and other potential technical approaches to automated discrimination** | Section 3 of this report signposts some of the key literature relating to DADM. It would be of further interest to better-understand how black box testing works in general, and whether there are other related technical approaches that could be used to prevent and detect discrimination. For instance, there has been significant focus on the notions of privacy, trust and |
| | [Computer science]* | trustworthiness – how could this understanding be widened to automated discrimination? |
| 5 | **Enriched understanding of statistical safeguards and data mining tools**<br><br>[Computer science + statistics + socio-cultural]* | It would be useful to interview/survey those directly involved with the design, development and employment of "black box" algorithms in order to better-understand the practicalities surrounding automated discrimination. For instance, the following questions could be asked to those directly involved with the design, development and employment of "black box" algorithms:<br>➢ To what extent are designers and developers aware about legal compliance?<br>➢ How do those directly involved currently prevent and detect instances of automated discrimination?<br>➢ What data mining tools (e.g. DADM) and other guidelines could assist those directly involved with the design, development and employment of "black box" algorithms to better safeguard against unlawful discrimination?<br>➢ What other safeguards are those directly involved already undertaking (e.g. privacy)?<br>➢ Do those directly involved consider automated discrimination to fall under the remit of trust?<br>➢ What practical measures are required to strengthen and raise awareness of statistical best practice? |

\* **The key expertise most likely required for a particular research recommendation.**

### 5.2.3 The development of an interdisciplinary approach (building on section 4)

Section 4 outlined the pragmatic need for an interdisciplinary approach to automated discrimination. Therefore, any future work would be focused on further development of this methodology, including both its quantitative and qualitative elements. Moreover, this would also involve developing a model that represents fairness and unfairness as explicit norms.

| | Recommendations | Brief overview |
|---|---|---|
| 1 | **Quantitative modelling**<br><br>[Interdisciplinary] | Need to examine factors such as: (1) statistical analysis of outputs and inputs; (2) the representation of fairness and unfairness as explicit norms; and (3) new assistive tools for automated discrimination prevention and discovery. |
| 2 | **Qualitative modelling**<br><br>[Interdisciplinary] | Need to examine factors such as: (1) the roles and responsibilities of stakeholders (stakeholder analysis); (2) internal training, policies and procedures (e.g. for legal compliance and statistical proficiency); (3) impact assessments; (4) the utility of certain supportive tools to assist with automated discrimination prevention and discovery; (5) due diligence; and, (6) mechanisms for accountability and transparency. |
| 3 | **Model testing and evaluation**<br><br>[Interdisciplinary] | Need to test and evaluate the model through scenarios, interviews/surveys and pilot testing (where applicable) with a wide range of stakeholders (e.g. software developers and legal experts). |

### 5.2.4 Summary – potential project deliverables in the short to long-term

In the short-term, the next step for this project would be to produce a publication based on this initial research to further raise-consciousness of the issues surrounding automated discrimination, and the potential interdisciplinary challenges ahead in its prevention and discovery.

In the medium- to long-term, there is significant scope for a larger scale interdisciplinary project that focuses on the research question: how, and to what extent, those directly involved with the design, development and employment of a specific "black box" algorithm can be certain that it is not unlawfully discriminating (directly and/or indirectly) against particular persons with protected characteristics (e.g. gender, race and ethnicity)? There is potential for a number of work packages to be defined, as illustrated by sections 5.2.2 and 5.2.3. The possible chief deliverables from such a project would be:

1. A common (interdisciplinary) understanding of automated discrimination both now and within the future digital environment;
2. The generation of data concerning stakeholder's perceptions of and attitudes to automated discrimination (e.g. through the use of interviews and surveys);
3. Publications and project reports linked to each work package;
4. Further consciousness-raising about the importance of statistical proficiency; and,
5. A new interdisciplinary approach to automated discrimination – "the model" – to include both quantitative and qualitative elements.